\documentclass[10pt]{iopart}

\usepackage{graphicx}
\usepackage[labelfont = bf,labelsep = period, font = footnotesize]{caption}
\usepackage{cite}
\usepackage{amstext}

\begin{document}
	\title[Two-color coherent control in photoemission from gold needle tips]{Two-color coherent control in photoemission from gold needle tips}
	\author{Philip Dienstbier\textsuperscript{1}, Timo Paschen\textsuperscript{1,\footnotemark} and Peter Hommelhoff\textsuperscript{1}}
	\address{$^1$ Department of Physics, Friedrich-Alexander-Universität Erlangen-Nürnberg (FAU), Staudtstraße 1, 91058 Erlangen, Germany, EU}
	\ead{philip.pd.dienstbier@studium.uni-erlangen.de}
	\begin{abstract}
	 	We demonstrate coherent control of photoemission from a gold needle tip using a two-color laser field. The relative phase between a fundamental field and its second harmonic imprints a strong modulation on the emitted photocurrent with up to 96.5\,\% contrast. The contrast as a function of the second harmonic intensity can be described by three interfering quantum pathways. Increasing the bias voltage applied to the tip reduces the maximum achievable contrast and modifies the weights of the involved pathways. Simulations based on the time-dependent Schrödinger equation reproduce the characteristic cooperative signal and its dependence on the second harmonic intensity, which further confirms the involvement of three emission pathways.
	\end{abstract}

\vspace{2pc}

 \footnotetext{Now with: Korrelative Mikroskopie und Materialdaten, Fraunhofer-Institut für Keramische Technologien und Systeme IKTS, Äu{\ss}ere Nürnberger Stra{\ss}e 62, 91301 Forchheim, Germany, EU} 
\maketitle

\section{Introduction}
Two-color laser fields formed by a strong fundamental laser pulse and its weak, phase-locked second harmonic became an important tool to manipulate and probe the electron emission dynamics in atomic and molecular gases. Complex and highly asymmetric waveforms can be generated by tuning the relative phase between both fields, changing their individual intensities or polarization angles. These sculpt fields can manipulate and probe the yield \cite{schumacher1994,muller1990}, angular structure \cite{yin1992,yin1995two,skruszewicz2015}, interference in momentum spectrum \cite{xie2012} and trajectories \cite{shafir2012} of emitted electrons with applications in the generation of high-harmonic \cite{mashiko2008} and terahertz \cite{dai2009} radiation. Metallic nanostructures, on the other hand, play an important technological role as DC and laser-triggered electron sources \cite{kruger2018}. Their usage ranges from high-resolution electron microscopes \cite{spence2013} and compact x-ray sources \cite{swanwick2014,graves2012} to recently developed on-chip, light-driven electronics \cite{krausz2014,rybka2016,bionta2020}. Thus, it is highly desirable to combine metallic solid-state systems with tailored electron emission in two-color laser fields.

So far, it has been shown that sweeping the relative phase between a fundamental field and its second harmonic modulates the current emitted from a tungsten needle tip with an unprecedented contrast of up to 97.5\,\% \cite{foerster2016,paschen2017,li2021}. The high contrast in this so called coherent control scheme applied to a tungsten needle tip can be well explained by the interference of two quantum pathways. Tungsten has many advantageous material properties such as the highest melting point among all metals, but it is a non-plasmonic material in contrast to, for example, gold. The creation of propagating surface plasmons and localized surface plasmons in the case of antenna structures, however, is often desired in light-matter interaction. The plasmon resonance helps to additionally increase optical near-fields at nanostructures \cite{dombi2020,thomas2013,thomas2015} and therefore reduces the required incident field strengths for non-linear photoemission. Propagating surface plasmons, on the other hand, can spatially separate the optical excitation from the electron emission as used in nanofocused electron point sources \cite{vogelsang2015,mueller2016}.

Here, we show that the coherent control scheme can be applied to the plasmonic material gold, too. A visibility of up to $96.5\%$ is obtained in the emitted photocurrent from a gold needle tip. In order to explain the scaling of the coherent signal with the second harmonic intensity, an additional, third pathway needs to be included. Simulations based on the time-dependent Schrödinger equation (TDSE) \cite{seiffert2018} reproduce the observed scaling and further validate the three-pathway model. Furthermore, we find that the achievable visibility can be suppressed by increasing the applied bias voltage \cite{paschen2017}, which is accommodated by changing contributions of the individual quantum pathways to the coherent signal.

\section{Experimental setup}
\begin{figure}[h!]
	\centering
	\includegraphics[width=0.75\linewidth]{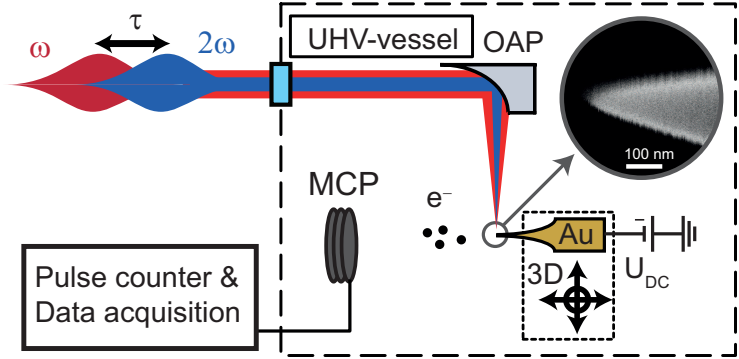}
	\caption{Experimental setup. Multi-cycle laser pulses centered at 1560 nm ($\omega$) and their second harmonic ($2\omega$) are collinearly focused onto the apex of a gold needle tip by an off-axis parabolic mirror within an UHV vessel. The emitted electrons are detected by a multi-channel plate (MCP) and counted as function of the optical delay $\tau$ between both laser fields. A bias voltage $\textrm{U}_\textrm{DC}$ is applied to the gold tip. Inset: Scanning electron micrograph of the gold tip with deduced apex radius of 25 nm and half-opening angle of $12^\circ$.}
	\label{fig:setup}
\end{figure}

The two-color field used in the experiment is composed of fundamental pulses centered around 1560\,nm with 74\,fs pulse duration and their second harmonics centered at $780\,$nm with $64\,$fs pulse duration. The fundamental pulses are delivered by an Erbium-doped fiber laser at a repetition rate of $100\,$MHz and the second harmonic is generated in a $100\,\mu$m thick $\beta$-barium borate crystal. Due to the parametric process the second harmonic has a fixed phase relation with respect to the fundamental field even if the carrier-envelope phase is not stabilized. A dichroic Mach-Zehnder type interferometer provides a tuneable optical delay between both fields, allows to adjust the individual intensities and to match the polarizations of both colors with the symmetry axis of the tip. A more detailed description of the optical setup is given in \cite{foerster2016}. The two-color field is tightly focused onto the apex of a gold needle tip inside a UHV vessel with a base pressure of $\approx 1 \times 10^{-10}\,$hPa as indicated in figure \ref{fig:setup}. The beam waists of the fundamental and second harmonic fields are $w_{\omega} = 4.0 \,\mu$m and $w_{2\omega} = 3.1 \,\mu$m ($1/e^2$ intensity radii). Electrons emitted from the tip are counted by a multi-channel plate (MCP) detector and the tip can be biased with a static voltage $\textrm{U}_\textrm{DC}$.

The gold tip was etched from a 0.1 mm thick poly-crystalline gold wire using a 90\% saturated aqueous potassium chloride solution \cite{eisele2011}. A tip apex radius of 25 nm and half-opening angle of $12^\circ$ can be estimated from a scanning electron micrograph (inset in figure \ref{fig:setup}). Nearfield enhancement factors of $FE_{\omega} = 6.5$ and $FE_{2\omega} = 5$ for the fundamental and second harmonic field are extracted from finite-difference time-domain simulations \cite{thomas2015} for the measured tip geometry. The maximum incident powers in this study are $P_\omega = 48\,$mW and $P_{2\omega} = 2.0\,$mW, which correspond to maximum near-field intensities of $I_{\omega} \approx 9 \times 10^{11} \textrm{ W}/\textrm{cm}^2$ and $I_{2\omega} \approx 4 \times 10^{10} \textrm{ W}/\textrm{cm}^2$. The work function for a clean gold surface $W = 5.4\,$eV \cite{kahn2016} results in minimal Keldysh parameters of $\gamma_\omega = 3.6$ and $\gamma_{2\omega} = 33$ placing our experiment clearly in the perturbative photoemission regime.

\begin{figure*}[h!]
	\centering
	\includegraphics[width=1\linewidth]{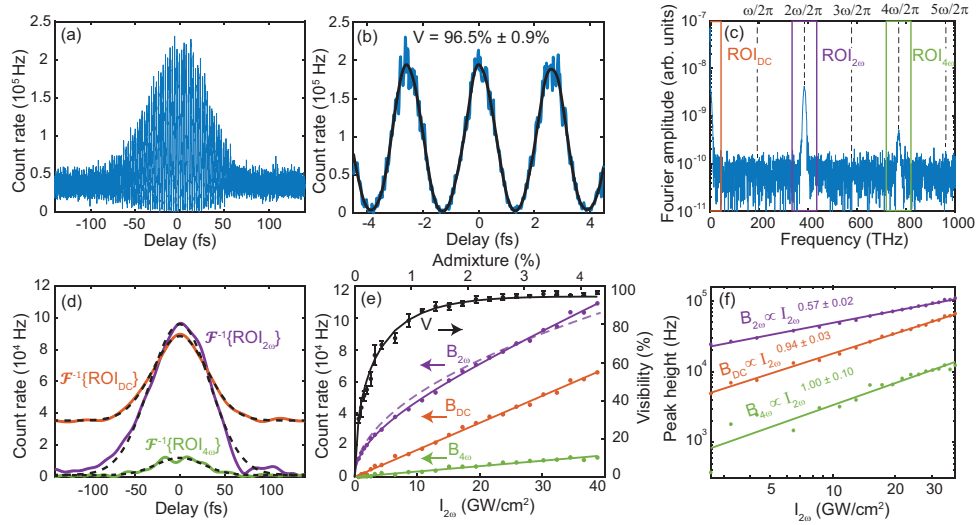}
	\caption{Coherent control of photocurrents from a gold needle tip. (a) Photocurrent as function of the delay between the $\omega$ and $2\omega$ field at intensities $I_{\omega} = 900 \textrm{ GW}/\textrm{cm}^2$ and $I_{2\omega}= 32 \, \textrm{GW}/\textrm{cm}^2$ with bias voltage $\textrm{U}_\textrm{DC} = -150 \, $V. (b) Count rate modulation with up to $96.5\% \pm 0.9 \%$ visibility in the center of the temporal overlap. The black line indicates a smoothing spline for the visibility estimation. (c) Amplitude of Fourier transform of the time domain data in (a). The cooperative signal is formed by Fourier components within regions of interest (ROIs) around the DC, $2\omega/(2\pi)$ and $4\omega/(2\pi)$ frequencies labeled $\textrm{ROI}_\textrm{DC}$, $\textrm{ROI}_{2\omega}$ and $\textrm{ROI}_{4\omega}$. (d) Inverse Fourier transform with additional Hilbert transform of $\textrm{ROI}_{2\omega}$ and $\textrm{ROI}_{4\omega}$. Gaussian fits according to equation (\ref{eq:Gauss}) are indicated by black dashed lines. (e) Peak heights $B_\textrm{DC}$, $B_{2\omega}$ and $B_{4\omega}$ of the Gaussian fits together with extracted visibility as function of the second harmonic intensity and admixture. Dashed line is a square root fit and solid lines are fits according to the three-pathway model. (f) Double-logarithmic representation of peak heights from (e) with power law exponents obtained by linear fits (solid lines).}
	\label{fig:power} 
\end{figure*}

\section{Coherent control at gold needle tips}
Changing the delay between the fundamental and second harmonic field near the center of the temporal overlap causes strong oscillations in the electron count rate as shown in figure \ref{fig:power}(a). 

The two-color field increases the count rate by more than five times or nearly fully suppresses the emission compared to separated pulses. By zooming into the center of the temporal overlap (figure \ref{fig:power}(b)) we observe an oscillation period of $T = 2.6$ fs  matching the period of the second harmonic field and determine a contrast of $96.5\% \pm 0.9 \%$. We use four local maxima and minima of a smoothing spline through the measured data to calculate the contrast or visibility $V$ and its standard deviation at the center of the time trace.
 
A Fourier analysis (figure \ref{fig:power}(c)) reveals that the time domain signal is composed of three characteristic frequency components centered around DC, $2\omega/(2\pi)$ and $4\omega/(2\pi)$ with $\omega= 192 \, \textnormal{THz} \times 2 \pi$ being the angular frequency of the fundamental field. To further identify the impact of the components on the time domain signal, we define regions of interest ($\textrm{ROI}_\textrm{DC}$, $\textrm{ROI}_{2\omega}$ and $\textrm{ROI}_{4\omega}$) around these Fourier components and transform them back individually. An additional Hilbert transformation provides their envelopes (figure \ref{fig:power}(d)), which can be well approximated by Gaussian functions 
\begin{equation}
	G_i = A_i + B_i \exp(-(\tau-\tau_{c,i})^2/D_i^2) \label{eq:Gauss}
\end{equation}
with offsets $A_i$, peak heights $B_i$, delay shifts $\tau_{c,i}$ and widths $D_i$ for $i = \left\{ \textnormal{DC},2\omega,4\omega \right\}$. As the peak visibility is obtained in the center of the trace, it only depends on the peak heights $B_i$ and the offset $A_\textnormal{DC}$.

 In figure \ref{fig:power}(e) we observe a steep rise in visibility as the second harmonic intensity $I_{2\omega}$ is increased. $B_\textrm{DC}$ and $B_{4\omega}$ depend linearly on $I_{2\omega}$. $B_{2\omega}$ deviates from an exact square-root scaling. Extracting the corresponding power law exponents in figure \ref{fig:power}(f) confirms this observation.

\begin{figure}[h]
	\centering
	\includegraphics[width=0.75\linewidth]{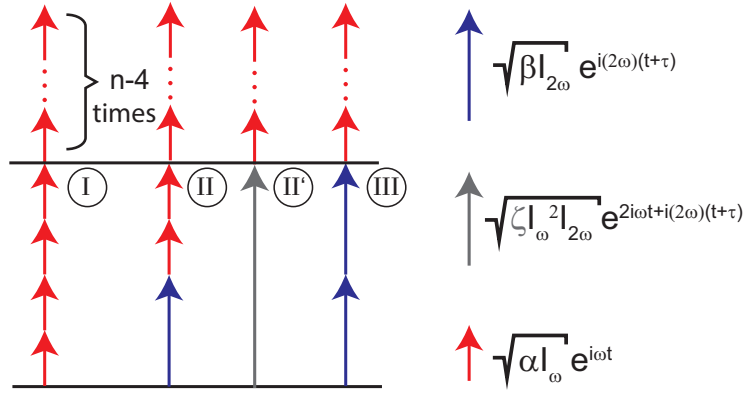}
	\caption{Quantum-pathway model. The pathway II (III) is created by exchanging two (four) photons of the fundamental field (red, weight $\alpha$) with one (two) photon of the second harmonic field (blue, weight $\beta$). The arrows depict partial amplitudes as indicated on the right. $\zeta^2 \neq \alpha^2 \beta$ is introduced as a new weight for pathway II labeled as pathway II'. }
	\label{fig:pathways}
\end{figure}

A model based on the interference between two quantum pathways is used to explain the experimental observations in the case of tungsten \cite{foerster2016}. The model considers the pathways I and II, illustrated in figure \ref{fig:pathways}, where two $(\omega)$-photons (red) are replaced by one $(2\omega)$-photon (blue) in pathway II. The weight $\alpha$ is related to the n-photon emission law $\alpha^n I_{\omega}^n$ and the weight $\beta$ is defined analogously for the second harmonic. The resulting (oscillating) transition probability is calculated by multiplying all partial amplitudes represented by the arrows in figure \ref{fig:pathways} within each path, adding the considered pathways and calculating the squared modulus to shift from amplitudes to rates.

Thus, the two-channel model predicts an oscillating term $(\alpha^{n-4} I_{\omega}^{n-4})$  $\left( 2 \alpha^3 I_{\omega}^3 \sqrt{\beta} \sqrt{I_{2\omega}} \right) \cos(2\omega\tau)$, whose prefactor is identified with $B_{2\omega}$. The resulting term $\alpha^n I_{\omega}^n$ corresponds to $A_{\textrm{DC}}$, where the small rate added by the second harmonic field is neglected in this model. The third term $(\alpha^{n-4} I_{\omega}^{n-4}) \times  \alpha^2 I_{\omega}^2 \beta I_{2\omega}$ is the incoherent background $B_{\textrm{DC}}$, which scales linear with $I_{2\omega}$. The two-pathway model cannot explain a frequency component at $4\omega/(2\pi)$ and expects $B_{2\omega}$ to follow an exact square-root dependence. 

High second harmonic admixtures make higher order substitutions likely. We now include pathway III, where four $(\omega)$-photons are replaced by two $(2\omega)$-photons. By evaluating all interference terms with the amplitudes given in figure \ref{fig:pathways} we obtain a component oscillating at $4\omega/(2\pi)$, which we identify as  $B_{4\omega}$, and the scaling laws
\begin{eqnarray}
B_\textrm{DC} &=  (\alpha^{n-4} I_{\omega}^{n-4}) \times \alpha^2 I_{\omega}^2 \beta I_{2\omega} = F_\textrm{DC} I_{2\omega} \label{eq:F_DC} \\
B_{2\omega} &= (\alpha^{n-4} I_{\omega}^{n-4})  \nonumber \\ 
&\times \left( 2 \alpha^3 I_{\omega}^3 \sqrt{\beta} \sqrt{I_{2\omega}} + 2 \alpha I_{\omega} \sqrt{\beta}^3 \sqrt{I_{2\omega}}^3 \right) \\
&= F_{2\omega,1} \sqrt{I_{2\omega}} + F_{2\omega,2} \sqrt{I_{2\omega}}^3   \\ 
B_{4\omega} &= (\alpha^{n-4} I_{\omega}^{n-4}) \times 2 \alpha^2 I_{\omega}^2 \beta I_{2\omega} = F_{4\omega} I_{2\omega}. \label{eq:F_4}
\end{eqnarray}

The three-channel model can explain the linear scalings of $B_\textrm{DC}$ and $B_{4\omega}$ as well as an increased order for $B_{2\omega}$. The corresponding fits match the data in figure \ref{fig:power}(e) and evaluating the fit coefficients $F_i$ results in

\begin{eqnarray*}
 F_\textrm{DC} &= (1.70 \pm 0.02) \times 10^{-6} \,\textnormal{Hz} \,\textnormal{cm}^2 \,\textnormal{W}^{-1} \\
 F_{2\omega,1} &= (4.47 \pm 0.06) \times 10^{-1} \,\textnormal{Hz} \,\textnormal{cm} \,\textnormal{W}^{-1/2} \\
 F_{2\omega,2} &= (2.7 \pm  0.2)\times 10^{-12} \,\textnormal{Hz} \,\textnormal{cm}^3\, \textnormal{W}^{-3/2} \\
 F_{4\omega} &=  (3.40 \pm 0.07) \times 10^{-7} \,\textnormal{Hz} \,\textnormal{cm}^2 \,\textnormal{W}^{-1}. 
\end{eqnarray*}

These four coefficients, however, cannot be retrieved by the simple relations for $\alpha$ and $\beta$. This can be seen by comparing $F_\textrm{DC}$ with $F_{4\omega}$, which do not fulfill $2F_\textrm{DC} = F_{4\omega}$ as expected from equations (\ref{eq:F_DC}) and (\ref{eq:F_4}). Allowing a new weight $\zeta$ for the exchange of one $(2\omega)$-photon with two $(\omega)$-photons (pathway II') we get 

\begin{eqnarray}
B_\textrm{DC} &=  (\alpha^{n-4} I_{\omega}^{n-4}) \times \zeta^2 I_{\omega}^2 I_{2\omega} \\
B_{2\omega} &= (\alpha^{n-4} I_{\omega}^{n-4}) \nonumber  \\
&\times \left( 2 \zeta I_{\omega} \alpha^2 I_{\omega}^2  \sqrt{I_{2\omega}} + 2 \zeta I_{\omega} \beta \sqrt{I_{2\omega}}^3 \right) \label{eq:comp}\\
B_{4\omega} &= (\alpha^{n-4} I_{\omega}^{n-4}) \times 2 \alpha^2 I_{\omega}^2 \beta I_{2\omega}.
\end{eqnarray}
Using $F_\textrm{DC}$, $F_{2\omega,1}$ and $F_{2\omega,2}$ we can evaluate

\begin{figure*}[h]
	\centering
	\includegraphics[width=1\linewidth]{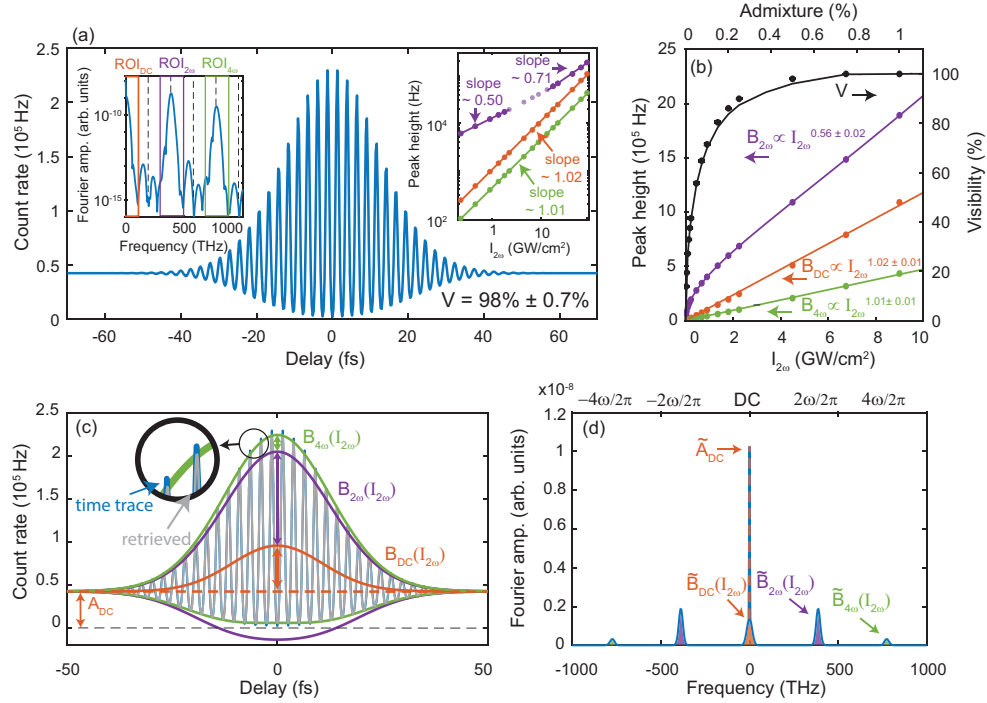}
	\caption{TDSE simulation of the cooperative signal. (a) Time domain representation of simulated photocurrent for pulse durations of $\tau_\omega = 25 \textrm{ fs}$ and $\tau_{2\omega} = 22 \textrm{ fs}$ and intensities of $I_{\omega} = 900 \,\textrm{GW}/\textrm{cm}^2$ and $I_{2\omega}= 4.5 \,\textrm{GW}/\textrm{cm}^2$.
		Left inset: Fourier amplitude and regions of interest $\textrm{ROI}_\textrm{DC}$, $\textrm{ROI}_{2\omega}$ and $\textrm{ROI}_{4\omega}$. Right inset: peak heights of Gaussian fits for the corresponding back transformed regions as function of the second harmonic intensity. The double-logarithmic representation clearly shows a change in the power law exponent for the peak height in $\textrm{ROI}_{2\omega}$. (b) Visibility together with peak heights $B_\textrm{DC}$, $B_{2\omega}$ and $B_{4\omega}$ (same as right inset of (a)) in linear scale as function of the second harmonic intensity and intensity admixture. Here, all data points are used to estimate the average power law exponent for $B_{2\omega}$. (c) Retrieved time domain signal (light gray) and original time domain signal form (a) by stepwise adding the Fourier components to the offset $A_\textrm{DC}$. For clarity only the envelopes are indicated. Small deviations are visible as the values for $B_\textrm{DC}$, $B_{2\omega}$ and $B_{4\omega}$ are taken from the fit curves in (b). (d) Fourier components with peak heights $\tilde{B}_\textrm{DC}$, $\tilde{B}_{2\omega}$ and $\tilde{B}_{4\omega}$ showing the corresponding frequency domain structure. The blue curve indicates the simulation.}
	\label{fig:tdse}
\end{figure*}

\begin{eqnarray*}
\alpha  &=  (8.70 \pm 0.05) \times 10^{-12} \,\textnormal{cm}^2 \,\textnormal{W}^{-1} \,\textnormal{Hz}^{-5} \\
\beta  &=  (  3.7 \pm  0.3) \times 10^{-10}  \,\textnormal{cm}^2 \,\textnormal{W}^{-1} \,\textnormal{Hz}^{-3}  \\
\zeta &=  (5.18 \pm    0.04) \times 10^{-16} \,\textnormal{cm}^3 \,\textnormal{W}^{-3/2} \,\textnormal{Hz}^{-13/2}.
\end{eqnarray*}

$F_\textrm{DC}$, $F_{2\omega}$ and $F_{4\omega}$ are by definition exactly reproduced, but also the independently retrieved value $F_{4\omega,r} = 2 \alpha^2 I_{\omega}^2 \beta =  (3.5 \pm  0.3) \times 10^{-7} \,\textnormal{Hz} \,\textnormal{cm}^2 \textnormal{W}^{-1}$ matches. The retrieved offset $A_\textrm{DC,r} \approx 2.9 \times 10^4$ Hz is also close to $A_\textrm{DC} \approx 3.5 \times 10^4$ Hz in the experiment. The maximum visibility in figure \ref{fig:power}(e) can be estimated using
\begin{eqnarray}
 V = \frac{B_{2\omega}}{A_\textrm{DC}+B_\textrm{DC}+B_{4\omega}}
\end{eqnarray}
and reproduces the experiment well, which is depicted by the solid black curve in figure \ref{fig:power}(e). We conclude that the third quantum pathway is needed in the presence of a frequency component around $(4\omega)/(2\pi)$ and allows us to accurately describe the experimental data.

\section{TDSE-simulations}
In this section we compare our experimental findings with model simulations based on the 1-dimensional time-dependent Schrödinger equation (TDSE), which is explained in detail in \cite{seiffert2018}. We use the same near-field intensities as in the experiment and assume a clean gold surface with work function $W=5.4$ eV and Fermi energy $E_\textnormal{F} = 5.2$ eV \cite{wu2020} without an applied bias voltage. In the experiment the effective barrier height is lower, as Schottky-lowering and possible residuals at the gold surface can reduce the work function \cite{kahn2016}. Further, we had to decrease the pulse durations to $\tau_\omega = 25 \textrm{ fs}$ and $\tau_{2\omega} = 22 \textrm{ fs}$ to keep the numerical effort manageable for a delay step size of 0.26 fs. Nevertheless, both pulses are clearly within the multi-cycle regime.

A typical simulated time trace is shown in figure \ref{fig:tdse}(a) and resembles the experiment closely. For a qualitative comparison we chose a trace with similar visibility and scaled the peak count rate to match the one from figure \ref{fig:power}(a). The Fourier amplitudes in the left inset of figure \ref{fig:tdse}(a) contain again the three characteristic components and show no additional features. The peak heights show a clear trend in the power law exponent of $B_{2\omega}$ exceeding the square root scaling for high second harmonic intensities as expected from the three-pathway model. $B_{2\omega}$ and $B_\textnormal{DC}$ maintain a linear scaling. Figure \ref{fig:tdse}(b) shows the same trends as in the experiment, even the average order of $B_{2\omega}$ coincides. The peak visibility in the simulation is reached for lower second harmonic admixtures, which is caused by the different effective barrier heights and uncertainties in the exact near field strengths in the experiment. The fit values

\begin{eqnarray*}
	F_\textrm{DC} &= (  1.18 \pm 0.01) \times 10^{-5} \,\textnormal{Hz} \,\textnormal{cm}^2 \,\textnormal{W}^{-1} \\
	F_{2\omega,1} &= (1.264 \pm 0.002) \,\textnormal{Hz} \,\textnormal{cm} \,\textnormal{W}^{-1/2} \\
	F_{2\omega,2} &= (8.08 \pm  0.03)\times 10^{-11} \,\textnormal{Hz} \,\textnormal{cm}^3 \,\textnormal{W}^{-3/2} \\
	F_{4\omega} &=  (4.72 \pm  0.03) \times 10^{-6} \,\textnormal{Hz} \,\textnormal{cm}^2 \,\textnormal{W}^{-1}
\end{eqnarray*}
allow again to estimate  

\begin{eqnarray*}
	\alpha &= (  4.931 \pm 0.007) \times 10^{-12}  \,\textnormal{cm}^2 \,\textnormal{W}^{-1}\, \textnormal{Hz}^{-7} \\
	\beta &= (1.26  \pm  0.02) \times 10^{-9} \,\textnormal{cm}^2\, \textnormal{W}^{-1} \,\textnormal{Hz}^{-4} \\
	\zeta &= ( 4.08 \pm  0.02)\times 10^{-16} \,\textnormal{cm}^3 \,\textnormal{W}^{-3/2} \,\textnormal{Hz}^{-9}.
\end{eqnarray*}

The obtained offset $A_{\textrm{DC}} = 4.25 \times 10^{4} \,\textnormal{Hz}$ slightly deviates from the retrieved value of $A_{\textrm{DC},r} =  3.4 \times 10^{4} \,\textnormal{Hz}$ as well as $F_{4\omega,r} =  (4.33 \pm  0.06) \times 10^{-6} \,\textnormal{Hz} \,\textnormal{cm}^2 \,\textnormal{W}^{-1}$. Furthermore, we can compare the values of $\alpha$ and $\beta$ against those estimated from an intensity-dependent yield-sweep providing $\alpha_p =   4.687  \times 10^{-12} \, \textnormal{cm}^2 \,\textnormal{W}^{-1} \,\textnormal{Hz}^{-7}$ and $\beta_p =  3.31 \times 10^{-11} \,\textnormal{cm}^2 \,\textnormal{W}^{-1} \,\textnormal{Hz}^{-4}$. The strong deviation in $\beta$ suggests that although all scalings are correctly predicted by the quantum-path model, the prefactors within the temporal overlap cannot be simply constructed from those of the individual fields.

In figure \ref{fig:tdse}(c) we investigate the formation of the time domain signal by adding up the corresponding regions of interest in the frequency domain in figure \ref{fig:tdse}(d) including their estimated widths. The main part of the signal is constructed by the offset $A_{\textrm{DC}}$ and $B_\textnormal{DC}$ defining the base line around which the (symmetric) oscillations with amplitude $B_{2\omega}$ appear. $B_{4\omega}$ flattens the trace for count rates close to zero and has to be included to prevent negative count rates. Adding up all these contributions gives nearly full agreement between retrieved and analyzed trace, even though $B_\textnormal{DC}$, $B_{2\omega}$ and $B_{4\omega}$ are obtained from the fit values for $\alpha$, $\beta$ and $\zeta$. In the Fourier domain, see figure \ref{fig:tdse}(d), this agreement is intuitively explained as the Fourier components are simply replaced by their fits and no complicated Fourier phase is present.

\section{Influence of the tip bias voltage}

\begin{figure*} [h]
	\centering
	\includegraphics[width=1\linewidth]{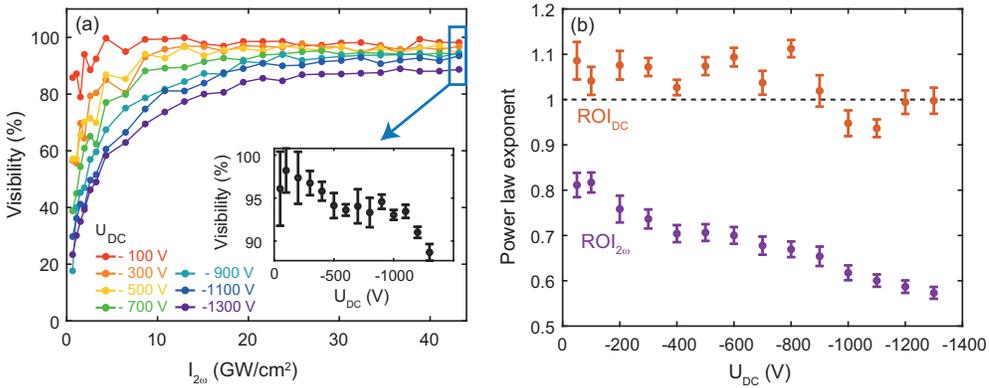}
	\caption{ Influence of the tip bias voltage. (a) Visibility as function of second harmonic intensity for bias voltages $\textrm{U}_\textrm{DC}$ from -100 V to -1300 V in -200 V steps at fundamental intensity $I_{\omega} = 540 \, \textrm{ GW}/\textrm{cm}^2$. Inset: visibility at the highest second harmonic intensity ($\approx$ maximal visibility). (b) Power law exponent for the DC- and $2\omega$ component as function of the bias voltage.}
	
	\label{fig:bias}
\end{figure*}

Gold tips with moderate apex radii ($>20$ nm) have high field emission thresholds ($<-1300$ V) while preserving sufficient photocurrents due to strong optical near-fields. Therefore, the big range of possible bias voltages is another important tuning knob in the coherent control scheme for gold. In figure \ref{fig:bias}(a) we investigate the influence of the bias voltage on the maximum achievable visibility and its dependence on the second harmonic intensity. Although all voltages are below the field emission threshold, a clear reduction of the maximum visibility is observed and the high visibility domain requires higher second harmonic intensities.  
The reduced visibility is accompanied by a shift in the power law exponent for the $2\omega$ component as shown in figure \ref{fig:bias}(b). The effective power law exponent is connected to the ratio of the components $2 \zeta I_{\omega} \alpha^2 I_{\omega}^2$ and $ 2 \zeta I_{\omega} \beta I_{2\omega}$, which scale differently with $I_{2\omega}$ (equation (\ref{eq:comp})). Their ratio is given by $(\alpha^2 I_{\omega}^2)/ \beta I_{2\omega}$ and shows that higher negative voltages further increase $\alpha I_\omega$ with respect to $\beta I_{2\omega}$. Thus, also the modifications of the visibility are most likely caused by changes in the weights of the respective quantum pathways.

 The bias voltage modifies the shape and height of the potential barrier simultaneously and therefore affects not only the values of $\alpha$,  $\beta$ and $\zeta$, but also their corresponding multi-photon orders. This interplay makes a model for the cooperative signal including bias voltages a challenging task, which is beyond the scope of this work.

\section{Conclusion and outlook}
In this manuscript we have shown that the coherent control scheme can be applied to gold needle tips and, as in the case of tungsten, a near 100$\%$ modulation of the photocurrent it possible. The quantum pathway model was extended by a third emission channel to describe the experiment and is supported by TDSE simulations. Applying bias voltages can modify the weights of the individual emission channels and leads to a reduced maximum visibility.

The requirement of independent weights for the three channels and the discrepancy in the extracted prefactors from those of independent pulses suggest modifications to the simple quantum pathway model. Analyzing the quantum pathway model using analytical solutions of the TDSE including bias voltages \cite{luo2019,luo2021} are therefore of highest interest for coherent control schemes in general and photoemission from metals in two-color fields in particular.

Using shorter and stronger fundamental pulses should in addition allow to observe trajectory modifications \cite{seiffert2018} driven by the two-color field and the spatial emission profile \cite{yanagisawa2009} could be modified by polarization shaped pulses. Thus, the combination of gold nanostructures with tailored emission in two-color fields may transfer electron wavepacket shaping capabilities previously reserved for atomic gases to nano-plasmonics, on-chip lightwave electronics and electron sources.

\section*{Acknowledgments}
This project was funded in part by the ERC grant ``Near Field Atto''and DFG SPP 1840 ``QUTIF''.

\section*{References}
\bibliography{references_two_color_gold}

\end{document}